\documentclass[12pt]{article}
\usepackage[dvips]{graphicx}
\usepackage{floatflt}
\usepackage{color}
\usepackage{colortbl}

\newcommand{\be}{\begin{equation}}
\newcommand{\ee}{\end{equation}}
\newcommand{\bean}{\begin{eqnarray}}
\newcommand{\eean}{\end{eqnarray}}
\newcommand{\bea}{\begin{eqnarray*}}
\newcommand{\eea}{\end{eqnarray*}}

\newcommand{\bc}{\begin{center}}
\newcommand{\ec}{\end{center}}

\suppressfloats[!]

\begin{document}

\title{MUON PAIR PRODUCTION IN PROTON-ANTIPROTON
            INTERACTIONS AT INTERMEDIATE ENERGIES.}


\author{ A.N.~Skachkova, N.B.~Skachkov,\\
Joint Institute for Nuclear Research,
Dubna, 141980, Russia\\
E-mail:Anna.Skachkova@cern.ch, skachkov@jinr.ru}

\maketitle

\begin{abstract}
\noindent 


      We use PYTHIA event generator to simulate the process 
    of muon pair production in  antiproton scattering upon
    the proton target at the energy of the antiproton equal to 
    $14GeV$, which may be one of the energies for the future 
    PANDA (GSI, Darmstadt) experiment operation.


\end{abstract}

\section{Introduction.}
\label{intro}



~~~~~Measurements of lepton pair production in hadron-hadron 
   interactions is of a big interest from the point of view of 
   study of the quark-parton structure of hadrons 
   \cite{Matv} and \cite{Drell}. 

     The best example is the discovery of J (also called as J/Psi)
   charmed meson
   confirmed later by
  e+e- experiment 
   tool to get out the information about the parton distribution 
   functions (PDF) in hadrons as it was already shown in a number 
   of high energy experiments \cite{DYexper} and 
   theoretical papers, devoted to the data  analysis in the 
   framework of QCD. 

     In this 
   connection it is worth to mention that up to now there is no 
   good theoretical understanding of the physics of this process.
   The best example of this situation is the wide use of such a
   "theoretical method"  of elimination of quite a noticeable 
   discrepancy among the lepton pair production data and the 
   theory predictions (quark-parton  model and its QCD extensions)
   as a simple multiplication of the theory predictions by the so 
   called phenomenological  "K-factors" which values vary in 
   the interval from 1.5 to 2.

      Intermediate energy experiments may play an important role as
   they allow to study the range where the  perturbative methods
   of QCD (pQCD) come into the interplay with a rich physics of bound 
   states.  The physics of hadron resonances formation and decays 
   is strongly connected with the confinement problem , i.e. with 
   the QCD dynamics at large distances. Its high precision and 
   detail experimental study would allow to make a serious
   discrimination between  a large variety of non-perturbative 
   approaches and models that are already proposed as well as 
   among those which are under development. 
  
      Therefore to define a boundary among perturbative and 
   non-perturbative approaches it may be instructive to make some
   estimations on kinematical distributions, connected with the 
   $\mu^{+}\mu^{-}$- or  $e^{+}e^{-}$- pairs
   production in the range of future PANDA experiment 
   energies 
    basing upon the well 
   known PYTHIA \cite{Sjost} event generator, which is widely and 
   successfully used in high energy experiments.  Such an 
   experience may allow to set the low energy limits for pQCD 
   application as well as it  may be also useful for comparison with the 
   predictions of the present (and future) non-perturbative 
   approaches and finding out their difference.


      It is worth to mention also that in any case the dominant
   mechanism of the lepton-anti-lepton pair production would be the 
   same in the most approaches because  in this special process
   $ p\bar{p} \rightarrow l\bar{l} + X $ 
   the  dominant contribution in any model would 
   come from the same quark pair annihilation  amplitude 
   ${q\bar{q} \to l\bar{l}}$.
   PYTHIA includes this amplitude as well
   as proper account of relativistic kinematics 

     To this reason, in present Note we shall simulate in framework of
   PYTHIA6 the scattering of the anti-protons with the energy 
   of $14 GeV$ on the proton target, taken as to be at the rest
   frame
   Thus the
   present simulation would not include  the effects connected 
   with the real detector conditions which would be the subject
   of the following papers.
%

     Thus,  the present simulation has the importance 
   as its own because it gives in some sense the "unbiased
   distributions" of particles as well as the corresponding 
   quark - parton distributions and allows, in principle,
   to estimate  the size  of the corrections needed to 
   reconstruct the original  input parton distributions. 
   Let us mention that in what follows  we choose 
   MRST4 \cite{MRST} parameterization of structure functions , 
   which  is one from the list of proton parton-distributions 
   set of PYTHIA6.  

     It should be noted also that the main part of this Note
   (Sections 1. and 2.)  can be easily applied to a case of
   electron-positron pair production. The main difference of 
   $e^{+}e^{-}$ case with that one of
   muon pair  production is that for $e^{+}e^{-}$ case 
   there would not a big  need of discussion of appearance
   of fake electrons in the same ``signal'' annihilation
   event due to negligibly small contribution of fake electrons
   from pions decays as comparing with their muon decay channel.    




%
\section{MUON DISTRIBUTIONS IN $P\bar P$  COLLISIONS.}

%


  ~~~ Here we present some distributions of muon physical 
   variables obtained by use of PYTHIA event generator. Its parameters
   were set to those values that allow fast simulation for the
   antiproton beam with the energy $E_{beam}$=14 GeV, which 
   corresponds to the center of mass frame energy of $p\bar p$ 
   collision $E_{cm}=5.3 GeV$.

     The muons produced in the corresponding hard QCD  $2\rightarrow 2$
   subprocess
${q\bar{q} \to \gamma^{*} \to \mu^{+}\mu^{-}}$ would be called in
    what follows as the "signal" ones, while those which will
   appear in event due to meson decays would be called as "decay"
   muons.




     The simulation was done for a case when both initial state
   radiation (ISR) and final one (FSR)  were switched on (i.e.
   with the following values of PYTHIA parameters MSTP(61)=
   MSTP(71)=1).
   The number of generated events was taken to be 100 000.

   The distributions of the signal muons energy $E^{\mu^{+/-}}$
   values   as well  as of  the modulus of the transverse
   momentum  $PT^{\mu^{+/-}}$
 and of  the polar angle
   $\theta^{\mu^{+/-}}$ (measured from the beam 
 direction)
   versus the number of generated events  are shown (from top
   to bottom, respectively) in Figure 1  of Section 6, which 
   contains all the Figures. The left  column in Figure 1 is for  
   $\mu^{-}$ distributions and the right one for $\mu^{+}$).
   There is no difference seen between  $\mu^{+}$  and  the 
   analogous e $\mu^{-}$ distributions in a case 
  when 
   the influence of magnetic field is not considered).

   One can see from the top raw of Figure 1 that the energy of
   muons may vary in the  interval $ 0 < E^{\mu} < 10GeV$ with
   a mean value $<E^{\mu}> = 2.6 $
 GeV and a peak at a 
   rather small value $E^{\mu}_{peak}$ =0.5 GeV. 
      
   The $PT^{\mu}$   spectrum , see middle raw of Figure 1, 
   has an analogous peak  $PT^{\mu}_{peak} = 0.5$ GeV/c.
   After this point both $E^{\mu}$  and $PT^{\mu}$
   spectra falls rather steeply but the spectrum of 
   $PT^{\mu}$ is confined in a more narrow interval 
   $0 < PT^{\mu} < 2$ GeV/c .

   The spectrum of events number (Nevent) versus the polar 
   angle  ${\theta^{\mu}}$ ( see bottom raw) has a peak at
   ${\theta^{\mu}} =10^{o} $
  and the mean 
   value  $ <\theta^{\mu}> =27.5^{o}$.

   From these plots we see that 
   while the most of signal muons are scattered into the 
   forward direction  $\theta^{\mu}<90^{o}$ still there is a
   small number of them that fly in the backward hemisphere
   ($\theta^{\mu} > 90^{o} $). We shall discuss this point in
   more
  details in what follows.

      From the view point of the background (the main
  source
   for it are the muons from charged pions and other 
   hadrons decays)  estimation it is usefull to have the
   set of plots analogouse to the  previouse  one, but  
   done separatley for the signal muons having the largest
   energy ( "fast" muons) in the muon pair produced in an
   event , and, 
   correspondently, for those having the smaller energy 
   ("slow" muon).  This set of plots for the signal muons
   are given in Figure 2 of the Appendix, where the left
   column is for  the  "slow" muon distributions and the right
   one is for "fast" muons.  


      We see from Figure 2 that  the energy spectrum (top raw) 
   of the fast 
  signal muons grows steeply from the point 
   $E^{\mu}_{fast}=0.5 GeV $
   (practically $99\%$ of fast muons have 
   $E^{\mu}_{fast} > 1 GeV $) to 
   the peak at the point $E^{\mu}_{fast}$ = $2.7 GeV$  and 
   then vanishes at  $ E^{\mu}_{fast}$ = $10 GeV $.

     In contrast to this picture the spectrum of less energetic 
   signal munons (left raw) has a peak at the point
   $E^{\mu}_{slow}$=0.5 GeV (where the spectrum of  fast
   muons only starts) and it practicaly vanishes at the point, 
   which corresponds to the mean value of the most energetic 
   signal muons, i.e. to $ < E^{\mu}_{fast} >$ = 3.9 GeV. Thus, one
   may say that the spectrum of slow muons in a pair is 
   very different from that of the fast ones.

     The difference between  the  $PT^{\mu}$ spectra 
   (see middle raw of  Figure 2) of the fast  and 
   slow  muons seams not to be so large and it
   results only in about 340 MeV shift to the left  of the peak
   position as well in the corresponding shift of the end point
   of slow muons spectrum. Both of these spectra demonstrait 
   that the most of  slow as well as of fast muons have
   $PT^{\mu} > 0.2$ GeV.

     This similarity of $PT^{\mu}$ spectra of fast and slow muons 
   results (due to a large difference of their energy spectra,
   i.e. of $P^{\mu}_{z}$ compronent) 
   in a large difference of their  polar angle ${\theta^{\mu}}$ 
   distributions (see bottom raw in Figure 2): ${\theta^{\mu}_{slow}}$
   spectrum is shifted to the right  as comparinring to that one for 
   ${\theta^{\mu}_{fast}}$ and their mean value
   $ < {\theta^{\mu}_{slow}} > $ =  $38.2^{o}$ is a more than 
   two times higher  than the analogouse mean value for the fast one:
   $ < {\theta^{\mu}_{fast}} > = 16.5^{o}$ .

     Another and the most important difference that is seen from 
   these angular 
 distributions is that all fast muons  fly
   in forward 
  direction 
   ${\theta^{\mu}_{fast}} < 90^{o}$ and their spectrum
   practically finishes at ${\theta^{\mu}_{fast}} = 60^{o}$,
   while about $17\%$ of slow muons  have 
   ${\theta^{\mu}_{slow}} > 60^{o}$  and  about of $5\%$ of them
   may  scatter into  the  back hemisphere.

\section{ SRUCTURE FUNCTIONS: \\ ~~~~u- AND d- QUARK DISTRIBUTIONS.}




    ~~~ Up to now we disscussed the results obtained without 
    any other kinematical cuts than those implemented 
    onto internal PYTHIA varaibles and needed to run 
    this generator at as low as possible values of beam energy.
    In our generation we have restricted the value  of the 
    invariant mass of $\mu^{+}\mu^{-}$ signal pair, produced
    in event. Namely we have taken the last parameter  
    $ M_{inv}^{\mu^{+}\mu^{-}}= {\hat{s}_{\;min}}$ to be restricted
    by the inequality  $ M_{inv}^{\mu^{+}\mu^{-}} >  1$ GeV.


    The distributions of Bjorken x-variables  are shown in
    Figure 3 
%
    for up- and antiup- quarks (top raw) and for
    down- and antidown- quarks (bottom raw) correspondingly.
    In what follows we shall refer to these distributions  
    as to the "unbiased" ones as the influence of different cuts
    onto muons energy $E^{\mu}_{cut}$ as well as the angle cuts,
    connected  with possible different geometry size of muon 
    muon system would be 
    studied latter. As it can be easialy seen from this plots
    the obtained distributions do not start from the point
    x=0, what take place for the used parton distribution
    parametrizations used in PYTHIA. Such a differnce at low 
    values of x appears due to the mention above cut on the
    value of muon pair invariant mass.

    The number of entries at these plots reflects the valence quark 
    flavour structure of the proton (the analoguse distributions 
    and the number of entries for the antiquarks show the absolute 
    similarity of quark and antiquarks distributions).

%
%
%
%
%
%

\section{  FAKE MUONS FROM MESON DECAYS.}



    ~~~ All what was said above may be to a good approximation
    applied also to a case of electron-positron pair production
    in the final state. The most difference of  $e^{+}e^{-}$ 
    case with that one of ${\mu^{+}\mu^{-}}$ pair production
    appears, as it was allready mentioned in the Introduction,
    when one turns to the problem of background or fake leptons
    in the same signal lepton pair production process.

      Really, the events with a pair of signal muons should
    contain  also some hadrons in the final state. The pions,
    produced  directly or in the  decays cascades of other
    hadrons,  may decay  in the  detector volume and thus
    serve as a source of background  muons that may fake the 
    signal ones produced in the annihilation subprocess.
    
    Figure 4 includes in the left column (as in the Figures 1 
    and 2)   
%
    the distributions (from top to bottom) of the  number of 
    events versus the energy $E_{PI}$= $E_{\pi}$,  the transverse
    momentum $ PT_{PI}$ = ${PT_{\pi}}$  and versus the polar angle  
    $TETA_{PI}$= $\theta_{\pi}$ of produced pions.

    The right hand column contains only one plot with the
    distribution of the total number NPI  of charged
    ${\pi}$-mesons in the signal events. One may see
    that there is a big number of events ( about$30\%$ )
    which do not  contain charged pions  at all  in the final 
    state  (they mostly have nucleon pairs in the 
    final state).
 
      This result is very important. Despite the  fact that 
    PYTHIA provides a good (at least one of the bests if not 
    the best one)
    but still a model approximation to the real hadronization
    effects, it indicates (in the absence of a complete physical
    and theoretical understanding of parton to hadrons fragmentation 
    processes) that it may happen so that about of one third 
    of events  with signal muon  pairs may appear practically  
    without additional  fake ( or internal background )
    decay muons content.
    That means that muon channel may be quite competitive to
    the   $e^{+}e^{-}$ one. In what follows we shall add more
    arguments in a favor of this suggestion.   
    
    From the same right hand plot we also see that  about  
    of  $25\%$ of events have only one charged  pion in the
    final state (appearing mostly in ${\Delta}$ resonance decays),
    and a bit less than  $25\%$ of events  have two charged pions.
    It demonstrates also  that about of $5\%$ of events 
    have 3 charged pions and there is only about of $1.5\%$ of
    events with 4 final state charged  pions.

    Figure 5 includes in the left column, respectively from
    top to bottom, the energy  $E^{\mu}_{dec}$,
    $PT^{\mu}_{dec}$  and  polar angle $\theta^{\mu}_{dec}$
    distributions 
%
    for  muons appearing from the discussed above charged 
    ${\pi}$ -meson decays. These distributions follow  
    the analogue spectra  of parent  pions, presented 
    in Figure 4.

%
%
       By comparing  these plots with those from signal muon 
    pairs shown in Figure 1, one may conclude that the 
    energy mean  value of the signal  muons 
    $< E^{mu} > $ =2.6 GeV does corresponds to the point in 
    the energy distribution of fake decay muons where the
    contribution of the decay muons is 
    already low. The same may be said about the PT-
    distribution of decay  muons. The mean value of  the PT-
    distribution of signal muons $PT^{mu}_{signal} = 0.7$ GeV
    (see Figure 1 ) does  corresponds to the point where the
    spectrum of $PT^{mu}_{decay}$ practically vanishes.
    
      The E- and PT- spectra of decay muons are a bit softer
    than those of  ``slow''  signal ones. They are  shifted by 
    300 MeV (i.e. by $30\%)$ to the left (see their mean values)  
    as comparing  with  those of signal slow muons. Analogously
    the fake decay muons polar angle  spectrum is 
    shifted also to the left by  $30\%$, i.e. by $14^{o}$.
    Still,  due to a big similarity of decay and ``slow''signal
    muons the former may appear as a serious background.
       
      Therefore  a reasonable cut upon the muons energy $E^{mu}$ 
    (as well as on the $PT^{mu}$) may lead to an essential
    reduction of the decay mesons contribution and allow to
    keep the main part of signal events.  
    Thus a cut $PT^{mu}> 0.2$ GeV may allow
    to get rid of a half of decay mesons  at the
  coast of 
    lost of $8\%$ of signal events .  The
    analogous strict cut $PT^{mu} < 0.4$GeV leaves more than $75\%$
    of signal muons and less than $15\%$ of decay mesons .
     
      Another way to discriminate the signal ``slow'' muons
    from the decay ones is to use the information about the
    position of the muon production  vertex.
                       
      The right hand side raw in Figure 5 contains (from top
    to bottom) the x-, y- and z-(i.e. along the beam direction)
    components of muon production vertex position, as given by
    PYTHIA simulation, i.e. in a space free of experimental
    setup around  the interaction point. 
    The distances at these distribution plots  are 
    given in millimeters. Thus one may see that the tails of
    the z-distribution may expand up to 100 meters, while those
    along x- and y- axis reach 40 meters distance. From this
    right hand plots  it is  clearly seen that in a case of 
    mentioned above  $66\%$ of  signal events that include 
    the decay  muons (i.e. addition to the signal muon pair, 
    produced in  annihilation subprocess) the size of the  
    detector (i.e., the ``decay volume'') may strongly reduce
    the number of charged  ${\pi}$-mesons which potentially 
    may produce muons in the decays because the most of
    parent pions would interact with  the detector
    components. Therefore one may expect that in the  real
    experimental  conditions the situation with the 
    contribution  of the  additional decay muons  
    may  become much easy. The detailed GEANT 
    simulation, which is in our nearest plans, should allow
    to get more definite predictions about the decay mesons
    contribution to the background.

\section{Conclusion.}
%

  ~~~~The energy, transverse momenta and polar angle (with respect
  to the beam  axis) distributions  of muons that may be
  produced in pairs in  $ p\bar{p} \rightarrow l\bar{l} + X $ 
  process , i.e. the  distributions of signal muons, are presented
  for a case of the proton target rest frame. These distributions
  were obtained by use of PYTHIA generator which parameters were
  adjusted to perform effective generation of physical events
  in a case of $\bar{p}$ beam energy equal to 14 GeV.

     It is shown that those of muons that are the most 
  energetic in a moun pair, i.e. fast muons, predominantly
  fly in forward direction (with the mean value
  $ < {\theta^{\mu}_{fast}} > = 16.5^{o}$) , while those which
  are less energetic in a  pair , i.e. slow muons, fly at larger
  angles (their mean value is 
  $ < {\theta^{\mu}_{slow}} > $ =  $38.2^{o}$).
  It is found that some of slow signal muons  may fly
  in backward hemisphere also. Thus, a good angle coverage 
  by muon system would be very usefull. 

    The analogous distributions for muons that appear from hadrons
  decays (mostly from pions), i.e. for decay muons, have shown that
  decay muons may fake the signal slow muons in the same events
  and thus, in principle, may be considered to appear as 
  a  serious bakground. Nevertheless, the distributions of the 
  number of generated events versus the number of charged pions in
  lepton pair production
  event has shown that, fortunatly, more than $30\%$ of signal events
  do not have  charged pions in the final state. On a top of this
  it seams to be clear from the obtained distribution of the space
  position of fake decay muon production vertex that a large amount 
  of these muons may be potentialy  rejected with the account of 
  the real detector decay volume. We plan as our next step to
  perform such type of complete GEANT  detector simulation with
  the analysed here events, generated by use of PYHTIA.  

    The auhtors are very gratefull to G.D.Alexeev for suggestion
  of this topic for study,  the interest to this work and multiple
  stimulating discussions of the questions concerned.






   \section{ Figures.}

\begin{center}
\begin{figure}[h]
\vskip -5mm
\hspace*{-15mm} \includegraphics[width=18cm,height=20cm]{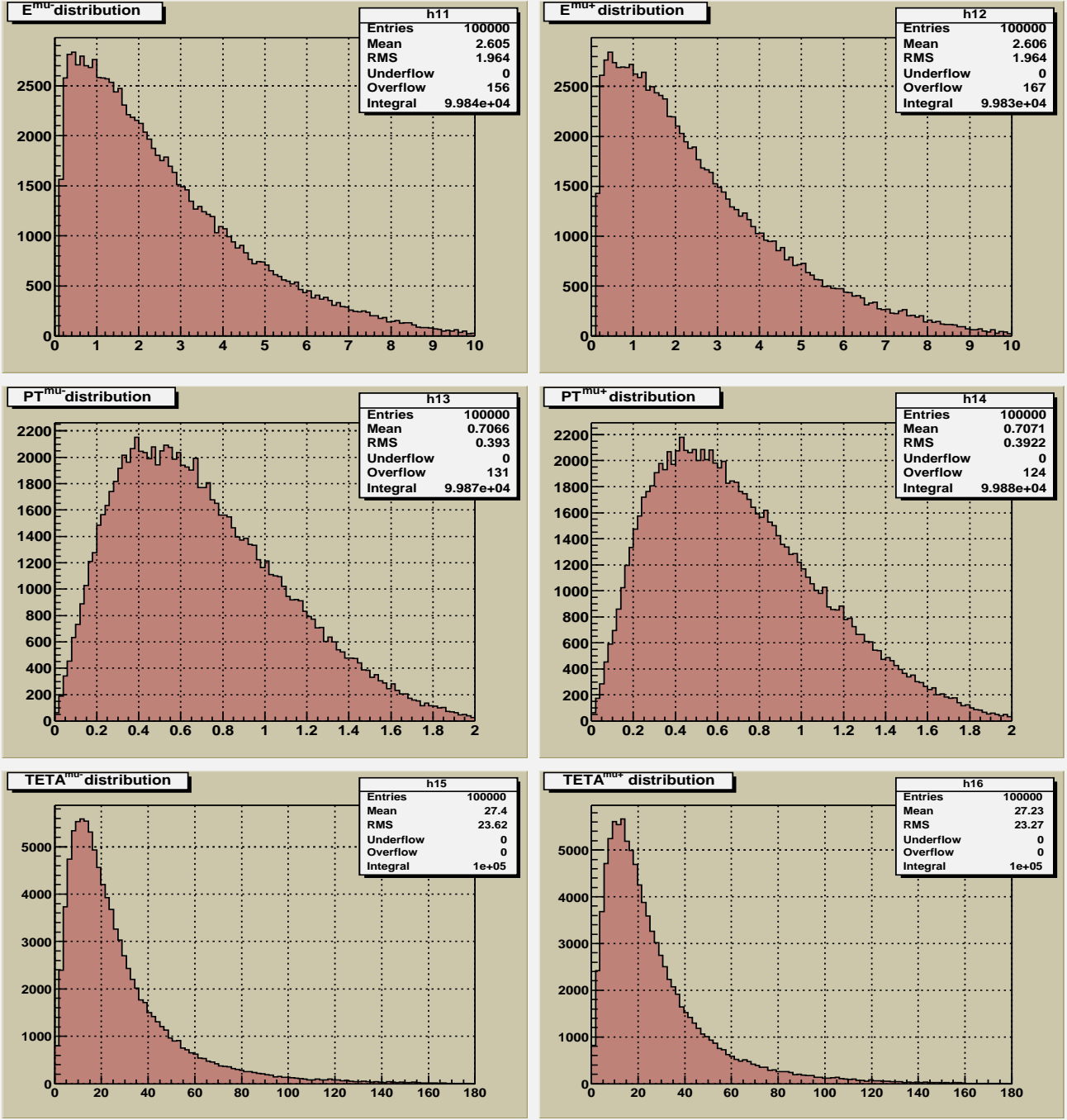}
\vskip 5mm
\caption{\hspace*{0.0cm}  Signal muons energy $E^{\mu^{+/-}}$ (top raw),
    the modulus of the transverse momentum  $PT^{\mu^{+/-}}$
    (middle raw) and  the  polar angle $\theta^{\mu^{+/-}}$
    (bottom raw)  distributions. Left column is for  $\mu^{-}$ 
    and right one for  $\mu^{+}$.}
\label{fig:1}
\end{figure}
\end{center}

\begin{center}
\begin{figure}[h]
\vskip -5mm
\hspace*{-15mm} \includegraphics[width=18cm,height=20cm]{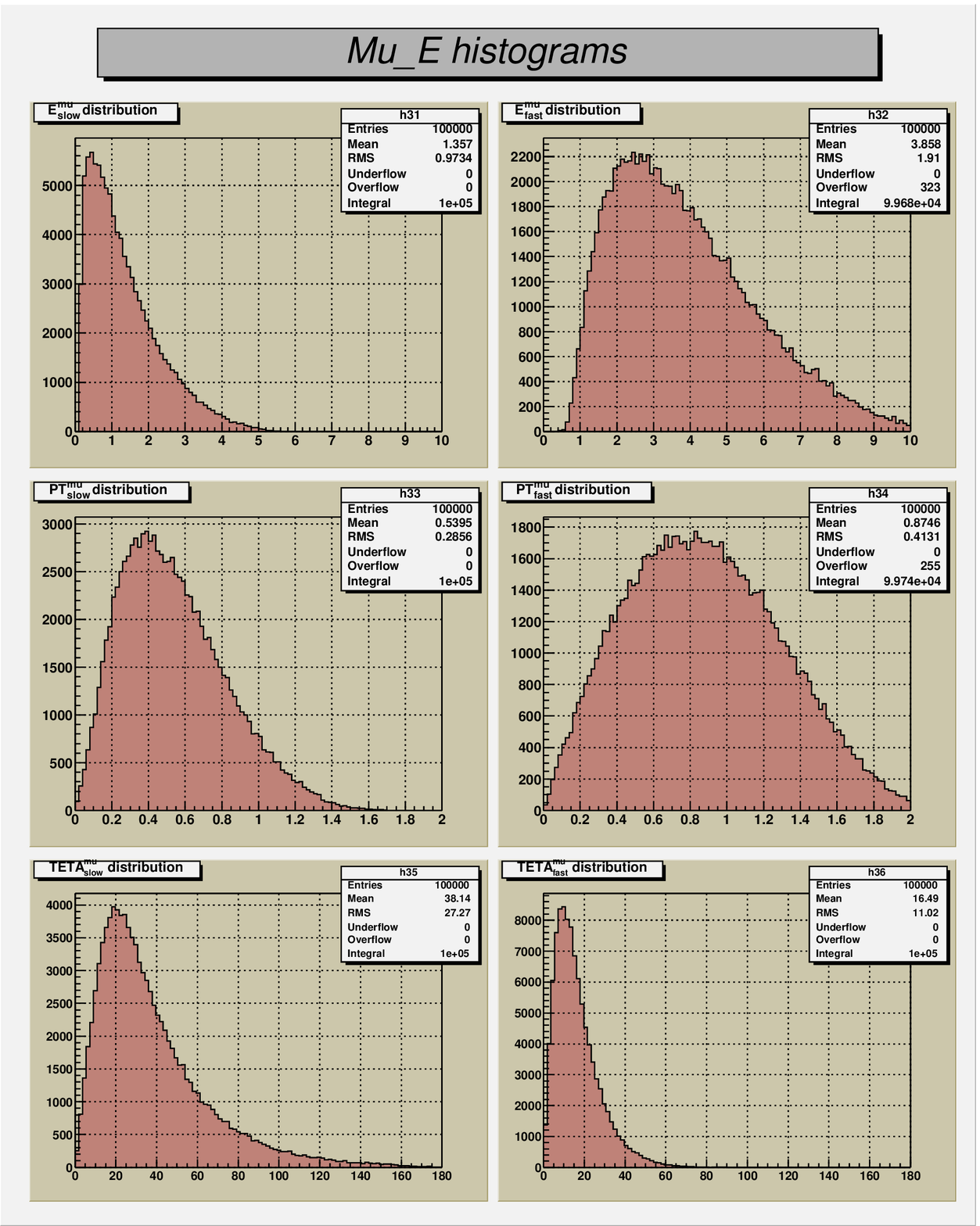}
\vskip 5mm
\caption{\hspace*{0.0cm} Signal muons distributions for muons with the
    largest energy ("fast" muons) in the muon pair 
    $E^{\mu}_{fast}$ (left column) and the smaller energy 
    ("slow" muons)  $E^{\mu}_{slow}$ (right column).
    Top raw includes their energies, in middle raw are PT and
    in bottom raw are $\theta^{\mu}$.}
\label{fig:2}
\end{figure}
\end{center}

\begin{center}
\begin{figure}[h]
\vskip -5mm
\hspace*{-15mm} \includegraphics[width=18cm,height=20cm]{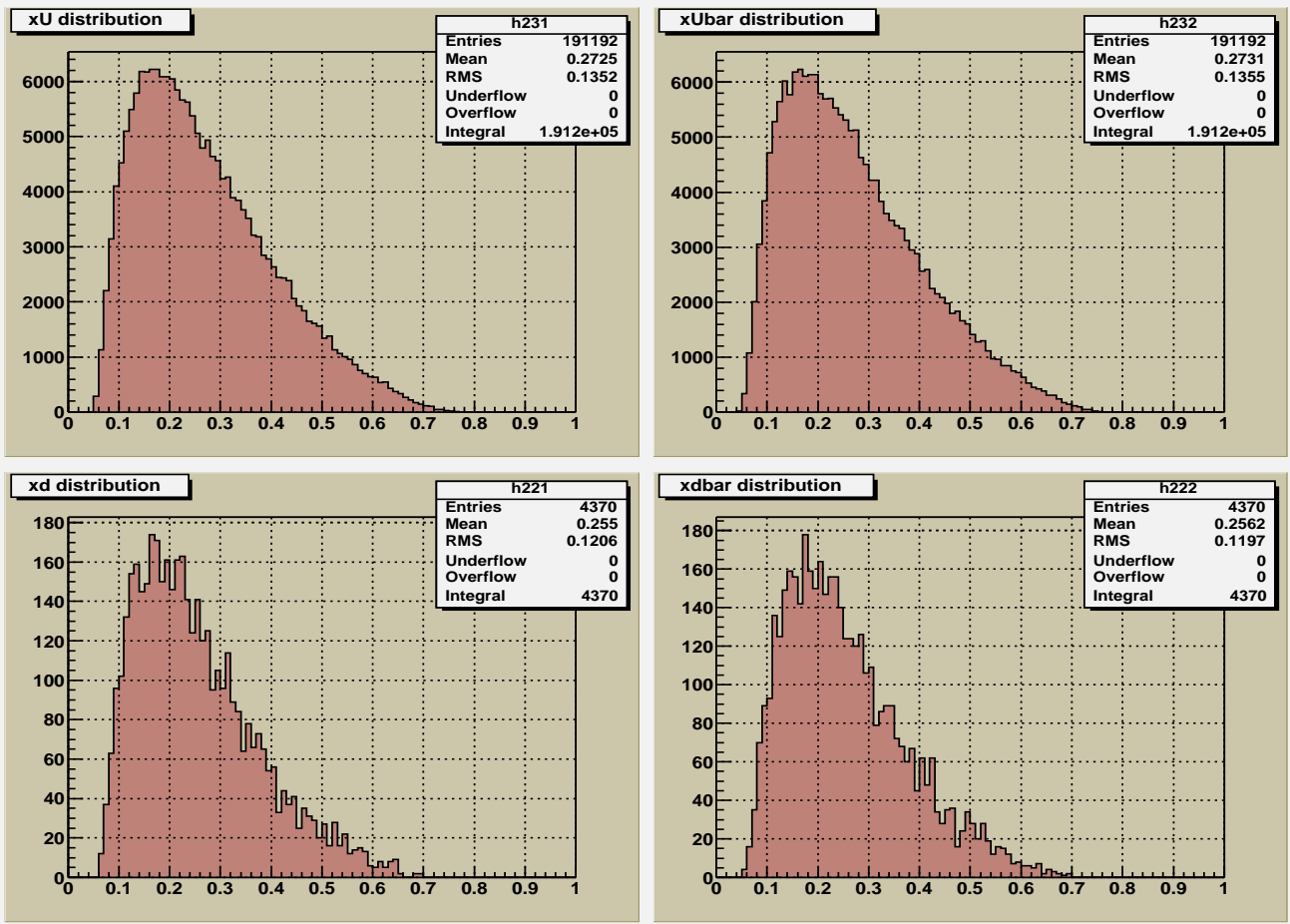}
\vskip 5mm
\caption{\hspace*{0.0cm}  Top raw shows the distributions of valence
              up -quarks and anti-quarks, while bottom raw 
              includes the analogous distributions of
              down-quarks and anti-quarks.}
\label{fig:3}
\end{figure}
\end{center}

\begin{center}
\begin{figure}[h]
\vskip -5mm
\hspace*{-15mm} \includegraphics[width=18cm,height=20cm]{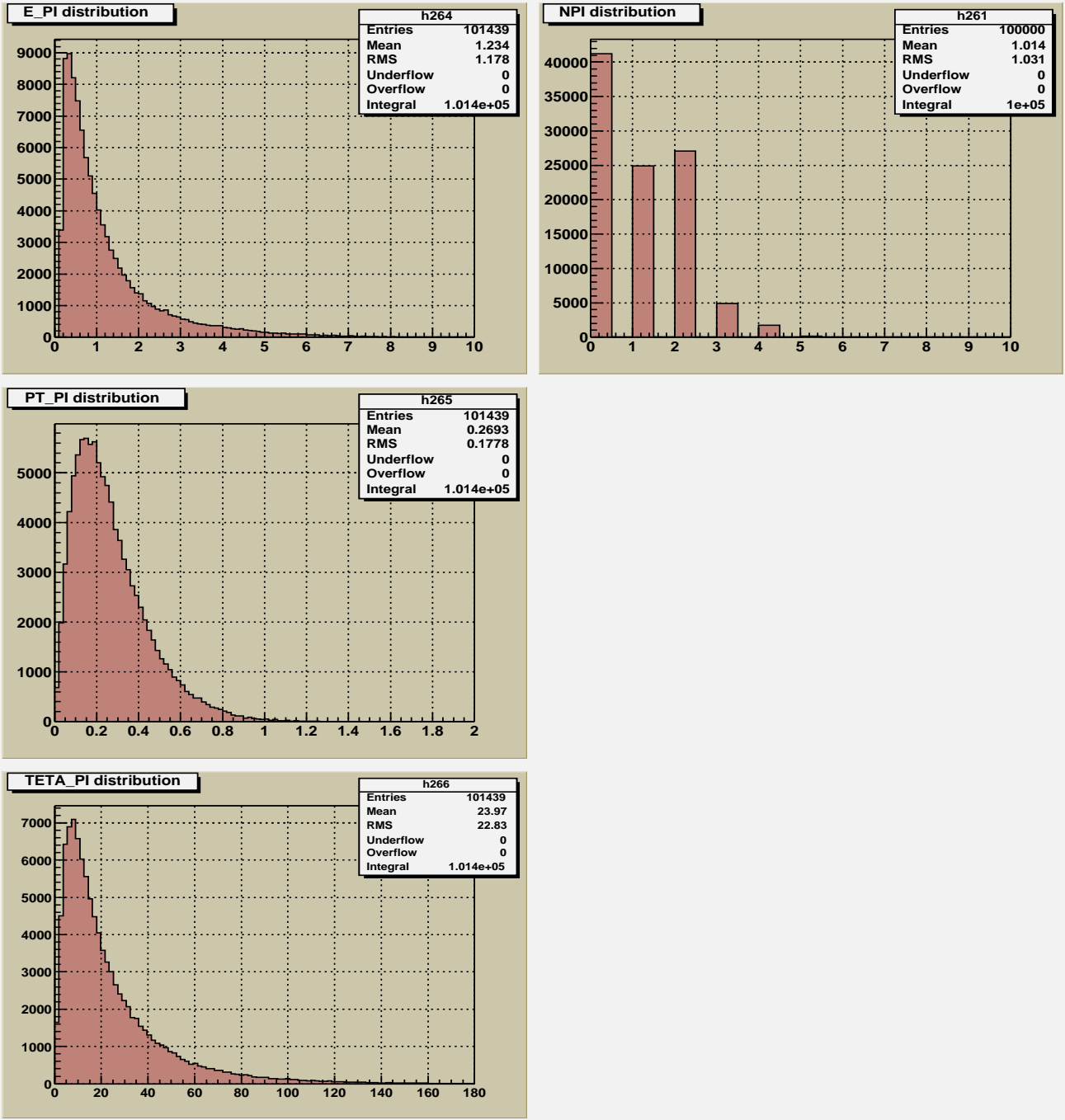}
\vskip 5mm
\caption{\hspace*{0.0cm} The left column  includes the distributions
              (from top to bottom, respectively) of number of
              events versus the energy $E_{PI}$=$E_{\pi}$, the
              transverse momentum  $PT_{PI}$= $PT_{\pi}$  and 
              versus the polar angle $TETA_{PI}$= ${\theta_{\pi}}$
              of produced pions. The right hand  plot shows the
              distribution of the total number (NPI)  of charged
               ${\pi}$-mesons in the signal events.}
\label{fig:4}
\end{figure}
\end{center}

\begin{center}
\begin{figure}[h]
\vskip -5mm
\hspace*{-15mm} \includegraphics[width=18cm,height=20cm]{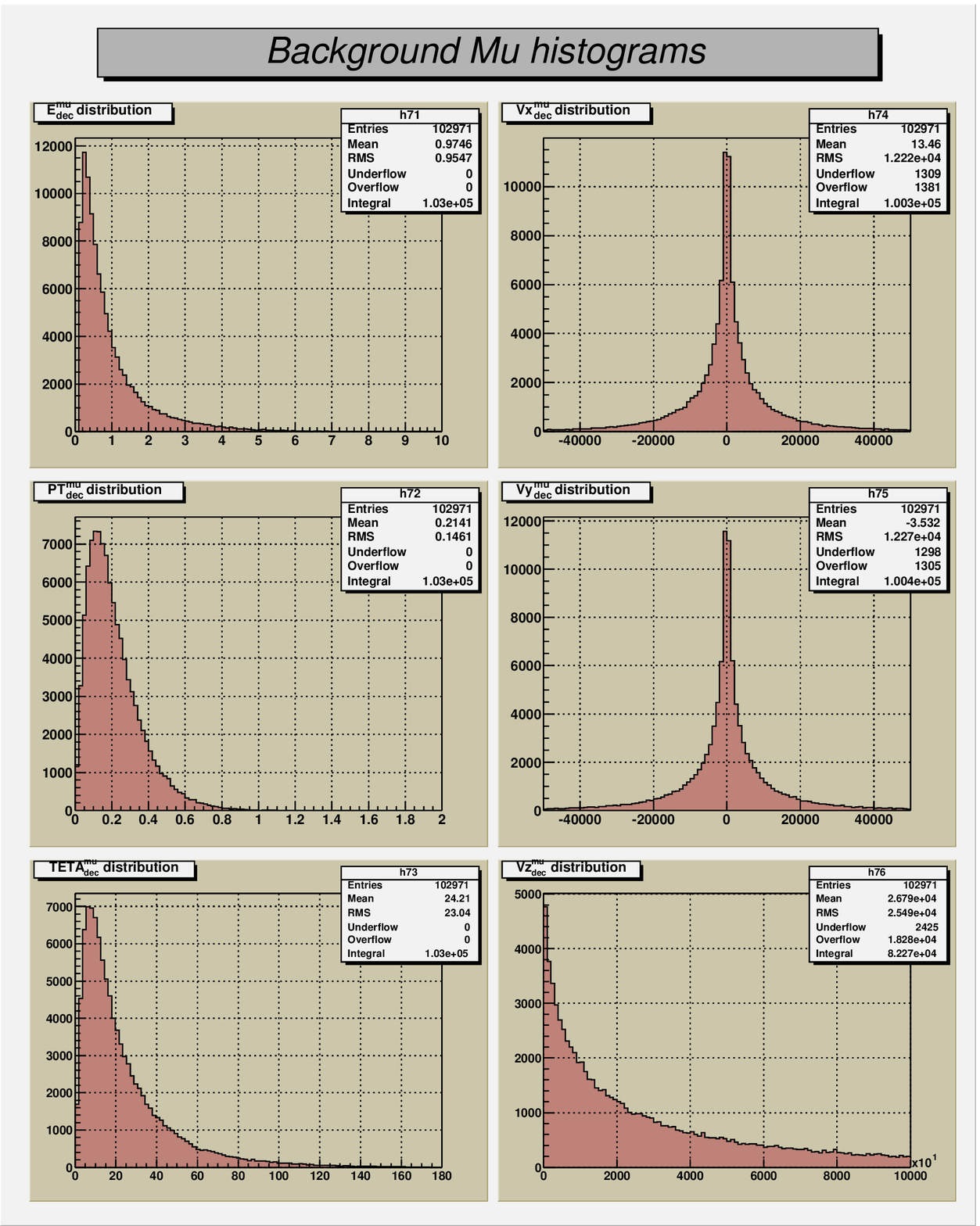}
\vskip 3mm
\caption{\hspace*{0.0cm} The left column  includes the distributions
              (from top to bottom, respectively) of number of
              events versus the energy $E_{PI}$=$E_{\pi}$, the
              transverse momentum $PT_{PI}$= $PT_{\pi}$  and 
              versus the polar angle $TETA_{PI}$= ${\theta_{\pi}}$
              of produced pions. The right column (from top to 
              bottom)  shows the distributions of  x-, y- and z-
              components of decay muon vertex position.}
\label{fig:5}
\end{figure}
\end{center}

\end{document}